\documentclass[acmlarge]{acmart}
\makeatletter
\def\@copyrightspace{\relax}
\makeatother
\renewcommand\footnotetextcopyrightpermission[1]{} 
\usepackage{wrapfig}
\usepackage{graphicx}
\usepackage{xcolor}
\usepackage{tabularx}
\usepackage{array}
\usepackage{xcolor}
\usepackage{longtable}
\usepackage{multirow}
\usepackage{caption}

\AtBeginDocument{%
  }

\setcopyright{none}
\copyrightyear{}
\acmYear{}
\acmDOI{}

\acmConference[]{}{}

\begin{document}

\title{My Machine and I: ChatGPT and the Future of Human-Machine Collaboration in Africa}

\author{Munachimso Blessing Oguine}
\email{munachioguine@gmail.com}
\affiliation{%
 \institution{National Open University of Nigeria}
 \city{Abuja}
 \state{Federal Capital Territory}
 \country{Nigeria}}

 \author{Chidera Godsfavour Oguine}
\email{oguinechidera7@gmail.com}
\affiliation{%
 \institution{University of Abuja}
 \city{Abuja}
 \state{Federal Capital Territory}
 \country{Nigeria}}

\author{Kanyifeechukwu Jane Oguine}
\email{janeoguine@gmail.com}
\affiliation{%
 \institution{University of Abuja}
 \city{Abuja}
 \state{Federal Capital Territory}
 \country{Nigeria}}

\renewcommand{\shortauthors}{Oguine, M. B. et al.}

\begin{abstract}
Recent advancements in technology have necessitated a paradigm shift in the people use 
technology necessitating a new research field called Human-Machine collaboration. ChatGPT, an Artificial intelligence (AI) assistive technology, has gained mainstream adoption and implementation in academia and industry; however, a lot is left unknown as to how this new 
technology holds for Human-Machine Collaboration in Africa. Our survey paper highlights to answer 
some of these questions. To understand the effectiveness of ChatGPT on human-machine 
collaboration we utilized reflexive thematic analysis to analyze (N= 51) articles between 2019 and 2023 obtained from our literature search. Our findings indicate the prevalence of ChatGPT for human-computer interaction within academic sectors such as education, and research; and trends also revealed the relatively high effectiveness of ChatGPT in improving human-machine collaboration.
\end{abstract}



\maketitle

\section{\textbf{Introduction}}
Over the last decade, technological efforts have been geared towards the optimization of human tech expertise and the automation of human capabilities through simulation; however;  huge gaps were identified in the proper simulation of human cognitive capabilities by Artificial intelligence (AI) technologies up until the recent breakthroughs in high computational power of computers, Deep neural networks and large language models (LLM).  

A nominal milestone in the advancement of AI has been the development of AI-assisted tools (such as ChatGPT,  etc.) a larger Language Model trained with Generative Pre-trained Transformer 3 (GPT-3)  developed by OpenAI to optimize open educational resources (OER) \cite{firat2023chat}. ChatGPT is an interactive chatbot, capable of using Natural Language Processing (NLP) to examine user input (usually text), and make an intuitive analysis of the text based on language patterns and relationships to produce contextual results relevant to the input text \cite{kalla2023study}. While this latest technological buzzword has gained prominence across several fields of scholarship, research, and industries, a whole lot is left to be understood regarding its current impact and future implications on users. 

Despite the successful implementation and positive reviews of ChatGPT reported by several scholars (\cite{kalla2023study}, \cite{biswas2023importance},\cite{liu2023summary}) the controversial significance and future of the AI assistive technology still poses a hot topic for discussion amongst domain experts and academics within Industry and Academia. The nuanced views and opinions on the rationale behind the efficiency of ChatGPT in processing tasks have posed several ethical and subjective questions, such as: (1) Are the results provided by ChatGPT biased or fair? (2) How accurate are the results provided by ChatGPT and can we trust them? (3) What ethical baselines do ChatGPT implement in structuring results or patterns, and how transparent are these baselines? (4) To what extent should ChatGPT be adopted? (5) What are the future implications of adopting ChatGPT across board? (6)Will ChatGPT necessitate human-machine collaboration or implement a holistic automated approach to problem-solving?
 
Grounded on global trends and the nuanced nature of questions and perceptions surrounding ChatGPT, this survey paper seeks to address some of the questions by providing a brief insight into the findings of recent studies with ChatGPT across several domains and implications. To do this, we conducted a literature review to explore the development of ChatGPT, the implications of Human-Machine collaboration, and several implementations of ChatGPT with human-machine collaborative objectives, and based on our findings we discuss future implications of ChatGPT across several domains and fields.

In line with the Computer-Supported Cooperative Work (CSCW) and SIG community’s interest in exploring the implication of Human-collaborative AI tools, this survey paper makes a survey contribution and also broadens the plethora of literary research works on chatGPT. This survey paper also provides a guide and potential research questions that can be explored by subsequent or future research.

\section{\textbf{Related Works}}
The emergence of chatGPT has shown a massive transformation and advancement of AI language technologies from Natural Language Processing to Natural Language Generation. Prior to the emergence and popularity of ChatGPT,  there have been existing conversational AI such as chatbots and virtual assistance some examples of which are Google Assistant (google), Siri(Apple), Alexa (Amazon), Cortana(Microsoft), and other web-based virtual assistance that provides personalized responses for users based on its natural language processing and user intents understanding capabilities.

ChatGPT is a generative conversational AI model released in 2022 by OpenAI, developed on a fine-tuned GPT 3.5 series model. The connotation of chatGPT can be described in two terms "chat" which gives reference to its function as a chatbot and "GPT" an abbreviation that stands for Generative rative Pre-trained Transformer; a type of Large Language Model (LLM).  ChatGPT is regarded as the "sibling model" of the InstructGPT model - developed on the concept GPT-3 language foundation model and prompt engineering. However, due to a major limitation of the GPT-3 language model which is its misalignment of user wants when performing natural language tasks, the pre-trained GPT-3 model was fined-tuned on data using supervised learning and the reinforcement learning from human feedback (RLHF) techniques to address the limitations of the GPT-3 model using human preferences as a reward signal to models to improve reliability and safety.

The OpenAI GPT 3.5 is a large language model that is an advancement of the GPT-3 model that is trained using deep learning-based approaches to provide detailed responses to instruction in a prompt through human-like text. In comparison to the GPT-3 model which trains on 175 Billion parameters, GPT-3.5 trains on a much larger dataset to give human-like feedback and generates the next responses from formerly existing prompts. Despite the improvements of the GPT foundation model to GPT-3.5, several limitations of ChatGPT have been observed such as Plausible-sounding but inaccurate and nonsensical responses, sensitivity to tweaks in prompt phrasing, language overuse, assumption of user intent given an ambiguous query due to lack of contextual understanding of input prompt, inability to identify and reject harmful or inappropriate user request, and behavioral bias, inability to understand human conversation nuances.

Recent advancements have been made to upscale the robustness and performance of deep learning and address certain limitations of GPT3.5 by OpenAI. The release of  GPT-4 has further extended the scope and capabilities of GPT models from just textual language models to multimodal models that can accept both image and textual prompts. Although not less capable than humans in performing real-world task, GPT-4 exhibit a high level of performance in several academic and professional benchmarks.

\subsection{\textbf{Human-Machine Collaboration}}
Human-machine collaboration also known as human-computer interaction, or human-machine cooperation, has emerged as a powerful paradigm in various domains and works of life. It leverages the unique collaborative strengths and capabilities of both humans and machines to enhance work outcomes, efficiency, and problem-solving capabilities. Human-machine collaboration is transforming the way tasks are accomplished, decisions are made, and problems are solved in various domains such as healthcare, finance, manufacturing, transportation, customer service, education, and more. By leveraging the unique strengths of humans and machines, organizations can harness the power of automation, data analytics, cognitive computing, and decision support systems to achieve higher levels of performance, efficiency, and innovation. Some of the domain applications of human-computer interaction are highlighted below:

\vspace{0.5cm}
\noindent
\begin{minipage}{15cm}
  \textbf{a. Education and Research:} Education and Research first witnessed the advancement in human-machine
  collaboration from algorithmic design and development to actual deployment of AI models. The adoption of AI in education has continuously impacted and shaped the educational system. A study by \cite{chen2020artificial} demonstrates how intelligent education systems through the use of adaptive learning techniques, personalized learning approach, computer vision, and facial recognition among several other techniques help in designing personalized learning for students, assessing teaching and learning methods, grading students performance, smart tutoring and hybrid learning. 
  
  \cite{freina2015literature}, shows how human-machine collaboration through virtual reality provides advantages such that it allows learners to explore physical objects that are not accessible in real life, to better understand and memorize them. \cite{wang2022collaborative} underscores the benefits of collaborative learning through human-machine collaboration, including enhanced problem-solving skills, critical thinking abilities, and teamwork. \cite{lee2021automated} demonstrate how automated grading systems can enhance grading consistency, efficiency, and timely feedback for students. \cite{smith2018data} highlights how human-machine collaboration in data analytics fosters evidence-based decision-making, leading to more effective instructional strategies and educational policies. By leveraging the strengths of humans and machines, education can be transformed to better meet the needs of individual students, enhance teaching practices, and drive educational success.

\end{minipage}

\vspace{0.5cm}
\noindent
\begin{minipage}{15cm}
  \textbf{b. Finance:} Technological advancements have permeated every sector of the economy, including the financial industry birthing the nomenclature Financial Technology (FinTech). One prominent application is the integration of Artificial Intelligence (AI) and Machine Learning (ML) systems, wherein humans collaborate with these technologies to enhance financial decision-making, risk assessments, risk management, fraud detection, and prevention, as well as investments and stock market predictions. Researchers are driven by the increasing industrial need for advanced risk management systems to explore sophisticated deep learning methodologies, such as Transfer Learning and Deep Reinforcement Learning, in the realm of financial risk management \cite{mashrur2020machine}.  Fraud detection is a major concern in the financial industry. 

The study by \cite{lee2021automated} focuses on human-machine collaboration for fraud detection using advanced analytics and artificial intelligence. Their research demonstrates that integrating human expertise with machine learning algorithms can effectively detect fraudulent activities, reducing financial losses and protecting the interests of financial institutions and customers. Monitoring changes in stock trends and making accurate stock market predictions are crucial for making investment decisions. \cite{ferreira2021artificial}, demonstrate how to realize huge financial investment profits by employing AI models, in stock market predictions.

\end{minipage}

\vspace{0.5cm}
\noindent
\begin{minipage}{15cm}
  \textbf{c. HealthCare:} Human-machine collaboration has gained enormous prominence in healthcare to improve patient care, enhance clinical decision-making, and streamline healthcare processes. This has been done by combining the knowledge and experience of healthcare professionals with the analytical power of machines to analyze vast amounts of patient data, biological and chemical data, including medical records, lab results, and imaging scans to assist healthcare professionals in diagnosing diseases and suggesting treatment options, and the discovery and development of new medications. AI systems are used to detect anomalies or changes in health conditions using advanced monitoring devices and wearable sensors to collect real-time patient data which health professionals can then review and make informed decisions and make timely interventions even in remote settings. AI algorithms can also assist in organizing and analyzing patient data, identifying patterns, and providing insights that can support clinical decision-making and population health management. Artificial intelligence (AI) is slowly transforming medical practice \cite{yu2018artificial}. Human-machine collaboration in healthcare holds great promise for improving patient outcomes, enhancing efficiency, and addressing healthcare challenges.
\end{minipage}

\vspace{0.5cm}
\noindent
\begin{minipage}{15cm}
  \textbf{d. Customer Services and Support:} Creating good customer service geared towards improving customer satisfaction is very crucial for any business or organization. Human-machine collaboration interventions like Virtual assistance and chatbots are examples that have proven to be very effective tools in improving the customer service industry. AI systems have become fully integrated into customer service by providing automated support and personalized responses for users. ML algorithms have been implemented in the analysis of data to generate useful insight and facilitate the decision-making process about customers and products or services.
\end{minipage}

\section{\textbf{ChatGPT in Human Machine collaboration}}
Despite some pushbacks, the implementation of ChatGPT across several domains and fields within industry and academia has gained broad adoption in actualizing the role of enhancing human-machine collaboration. To gain an in-depth understanding of our research topic, we adopted a systemic literature review process proposed by \cite{oguine2023investigating}, by conducting a literature search on Google Scholar aimed to synthesize and aggregate existing literary work at the intersection of the keywords; “chatGPT” and “Human-machine collaboration”. For our search, we scoped down the initial 66 results to articles published between 2019 and 2023. To ensure the relevancy of the articles to our research objectives, we filtered the articles by reading through the titles and abstracts, excluding those not directly relevant to our research topic.  After these processes, we obtained a total of 51 articles relevant to our research topic.

To gain an in-depth understanding of the role of chatGPT in Human-Machine collaboration, we utilized both inductive and deductive reflexive thematic analysis to obtain patterns through themes from the 51 articles. To achieve this we utilized qualitative coding to obtain codes from each article, after which we grouped the codes based on related patterns under groups. Table 1 shows a codebook containing the results of our search.

\begin{table*}[h]
    \caption{Codebook Dimensions, Themes, and Codes}
    \label{tab:my_label}
    \centering
    \begin{tabularx}{\textwidth}{|X|X|X|}            
        \hline
        \textbf{Dimension} & \textbf{Code} & \textbf{Definition} \\
        \hline
        Areas of Human-Machine Collaboration & Research (63\%, N=32) & Articles that highlight implications of Human-machine collaboration in Research \\
        \cline{2-3}
        & Education (18\%, N=9) & Articles that highlight implications of Human-machine collaboration in Education\\
        \cline{2-3}
        & Manufacturing (7\%, N=4) & Articles that highlight implications of Human-machine collaboration in Manufacturing\\
        \cline{2-3}
        & Health (6\%, N=3) & Articles that highlight implications of Human-machine collaboration in Health \\
        \cline{2-3}
        & Artistry and Creativity (4\%, N=2) & Articles that highlight implications of Human-machine collaboration in Artistry and Creativity \\
        \cline{2-3}
        & Military (2\%, N=1) & Articles that highlight implications of Human-machine collaboration in Military \\
        \hline
        Sectors of Human-Machine Collaboration & Academia (63\%, N=32) & Articles showing the adoption of human-machine collaboration in academia \\
        \cline{2-3}
        & Industry (37\%, N=19) & Articles showing the adoption of human-machine collaboration in Industry \\
        \hline
        ChatGPT Effectiveness & Highly Effective (33\%, N=17) & Articles showing the effectiveness of ChatGPT in promoting Human-Machine collaboration \\
        \cline{2-3}
        & Moderately Effective (31\%, N=16) & Articles showing the moderate effectiveness of ChatGPT in promoting Human-Machine collaboration \\
        \cline{2-3}
        & Lowly Effective (30\%, N=15) & Articles showing the low effectiveness of ChatGPT in promoting Human-Machine collaboration \\
        \cline{2-3}
        & Not Effective (6\%, N=3) & Articles showing the ineffectiveness of ChatGPT in promoting Human-Machine collaboration \\
        \hline
    \end{tabularx}
\end{table*}

\subsection{\textbf{Human-Machine Collaboration:}}
Qualitative coding was used to classify the areas of Human-Machine collaboration implementing ChatGPT. It can be observed from the codebook above that chatGPt has gained wide adoption in enabling human-machine collaboration within the field of research as indicated by 63\% (N=32) of the articles found, we observed that another area gaining prominence in the adoption of chatGPT for human-machine collaboration is the field of education with 18\% (N=9) of articles. Other areas that have seen the adoption of chatGPT based on the article search include manufacturing 7\% (N=4), health 6\% (N=3), artistry and creativity 4\% (N=2) and the military 2\% (N=1). Based on our findings, there is a likely tendency of chatGPT to gain wider implementation across certain areas of human-machine collaboration than others.

\subsection{\textbf{Sectors of Human-Machine Collaboration:}}
The disparity in opinion on the implications of human-machine collaboration to either industry or academic sectors according to \cite{lee2014service}, has recently become a baseline to approve or disapprove the adoption of human-machine collaboration in recent times. Based on this, we pre-established two themes (either Academia or Industry) to categorize the papers obtained from our literature search. This is to establish the application and implication of chatGPT in enabling human collaboration within the two sectors. From our codebook, it can be observed that ChatGPT has gained with adoption and utilization in the academic sector with 63\% (N=32) of the articles suggesting either its relevance or utilization. The remaining papers from the search 37\% (N=19) highlighted the role of ChatGPT in human-machine collaboration within the industrial sector. 
Overall, a noticeable trend that can be inferred is that chatGPT is more prominent at this time within academia than in Industry sectors.

\subsection{\textbf{The Effectiveness of ChatGPT in Human-Machine Collaboration:}}
One question this survey paper tries to answer is the effectiveness of ChatGPT in facilitating or impeding human-machine collaboration. To do this we pre-established four (4) themes to categorize the effectiveness of ChatGPT asserted by each article on human-machine collaboration. From our codebook (Table 1), 33\% (N=17) of the articles highlighted a highly positive implication of chatGPT in improving human-machine collaboration. 31\% (N=16) of the articles suggested a moderate effectiveness of chatGPT in supporting human-machine collaboration across board, 30\% (N=15) highlighted a low effectiveness of chatGPT in facilitating human-machine collaboration and finally, it was observed that 6\% (N=3) of the articles showed no effectiveness of chatGPT in facilitating or impacting human-computer collaboration.
Based on the findings from our thematic analysis, it can be concluded that current articles suggest the effectiveness of chatGPT in facilitating human-machine collaboration, however, there are still some factors to be considered to ensure near-perfect human-machine collaboration using ChatGPT.

\section{\textbf{The future of Human-Machine Collaboration in Africa}}
The landscape of human-machine collaboration continuously undergoes a transformative shift as advances in AI revolutionize human interactions and collaborations with machines. Traditional human-to-human collaboration is being replaced by the integration of AI across diverse domains like healthcare, finance, education, politics, entertainment, and journalism, as AI technologies become ubiquitous. The pervasive presence of AI systems empowers individuals and industries to forge innovative partnerships with intelligent machines, unlocking limitless possibilities for enhanced productivity, efficiency, and problem-solving capabilities. This widespread adoption of AI brings forth tremendous benefits for human-related tasks while presenting a multitude of opportunities and challenges that demand exploration and consideration from all technological stakeholders. Embracing this era of ubiquitous AI necessitates a thoughtful balance, where the benefits of collaboration are reaped while effectively addressing the challenges that arise.

Recent developments in NLP, like ChatGPT and subsequently GPT-4, have led to the creation of generative conversational AI. With a broader general knowledge base, higher reasoning, and problem-solving abilities, the GPT-4 model is capable of simulating human-like conversations and generating solutions to difficult problems. ChatGPT has demonstrated immense benefits in terms of its relevance to human-machine collaboration. It acts as a recommender system, a writing assistant for scholarly and professional articles, a virtual and personal assistant, a programmer, and an alternative search engine. By integrating NLP technologies with other AI technologies like computer vision, deep learning, and sensor fusion, there is a potential to enhance the robustness and accuracy of intelligent and autonomous vehicle performance. According to \cite{lemons2022stanford},  AI technologies when incorporated in the disability space, will be very useful for teachers and kids in the design of creative instructional material and development of personalized smarter tutoring that is responsive to student needs. 

As AI technologies gear towards transformative advancements, they will play an increasingly prominent role in supporting and enhancing the future of human-machine collaboration in decision-making processes. The integration of trustworthy AI algorithms and machine learning technologies can provide valuable insights, data analysis, and predictive modeling to inform decision-making across various industries and sectors. This collaborative approach can lead to more informed and accurate decision-making, improved efficiency, and optimized outcomes through the combination of human expertise, intuition, and contextual understanding with the computational power and analytical capabilities of AI systems.

Several concerns have been raised by policymakers, parents, educational institutions, scholars, and industry experts regarding the potential risk of AI in society, despite its benefits to productivity. Ethical concerns, and privacy issues regarding the access, use, and protection of data collected by AI systems. Over-reliance on AI-generated insights without proper human oversight can produce flawed decision-making and reinforce bias. Transparency,  accountability, and trustworthiness concerns AI decision making especially in critical areas like healthcare and finance. According to OpenAI, GPT-4 will continue to see more improvements in its algorithms to accommodate more user needs,  similarly, other AI systems will undergo major improvements. This poses a potential risk of over-dependence on AI in daily activities, which could in turn lead to a loss of creativity and, a reduction in human expertise, skills, and ability to make independent, autonomous decisions without the help of AI systems. 

\section{\textbf{SUMMARY AND CONCLUSION}}
In summary, our survey paper highlights the nuanced implications, possibilities, and capabilities of chatGPT on the current and future human-machine collaborative endeavor in Africa. Firstly, we discussed the aims, motivations, and objectives of the survey paper, which is aimed at exploring the nuanced dynamics and controversies surrounding the adoption of chatGPT for human-machine collaborative tasks. Next, to gain an in-depth understanding of chatGPT, we discussed briefly the fundamental details of the development of chatGPT by describing the developmental processes. Our survey paper also explores the recent literature work on human-computer collaboration across several fields (such as health, education, etc.). Our survey paper also makes an empirical research contribution by conducting reflexive thematic analysis on search results obtained at the intersection of chatGPT and Human-Machine Collaboration. Our findings indicated a prevalence of chatGPT in human-machine collaboration with a larger focus on the research field. We also discovered the nuanced effectiveness of chatGPT in facilitating human-machine collaboration. While notable efforts have been observed from articles indicating the possibility of chatGPT in promoting human-machine collaboration, there are certain factors that may impede this possibility. Finally, we discussed our opinion on the future of human-machine collaboration based on our findings from our research.

\bibliographystyle{acm}
\bibliography{samples/refer}

\begin{thebibliography}{10}

\bibitem{biswas2023importance}
{\sc Biswas, S.}
\newblock Importance of chat gpt in agriculture: According to chat gpt.
\newblock Available at SSRN 4405391.

\bibitem{chen2020artificial}
{\sc Chen, L., Chen, P., and Lin, Z.}
\newblock Artificial intelligence in education: A review.
\newblock {\em IEEE Access 8\/} (2020), 75264--75278.

\bibitem{ferreira2021artificial}
{\sc Ferreira, F. G. D.~C., Gandomi, A.~H., and Cardoso, R. T.~N.}
\newblock Artificial intelligence applied to stock market trading: A review.
\newblock {\em IEEE Access 9\/} (2021), 30898--30917.

\bibitem{firat2023chat}
{\sc Firat, M.}
\newblock How chat gpt can transform autodidactic experiences and open
  education.
\newblock Department of Distance Education, Open Education Faculty, Anadolu
  Unive.

\bibitem{freina2015literature}
{\sc Freina, L., and Ott, M.}
\newblock A literature review on immersive virtual reality in education: State
  of the art and perspectives.
\newblock In {\em 11th International Conference eLearning and Software for
  Education\/} (2015).

\bibitem{kalla2023study}
{\sc Kalla, D., and Smith, N.}
\newblock Study and analysis of chat gpt and its impact on different fields of
  study.
\newblock {\em International Journal of Innovative Science and Research
  Technology 8}, 3 (2023).

\bibitem{lee2014service}
{\sc Lee, J., Kao, H.~A., and Yang, S.}
\newblock Service innovation and smart analytics for industry 4.0 and big data
  environment.
\newblock {\em Procedia cirp 16\/} (2014), 3--8.

\bibitem{lee2021automated}
{\sc Lee, S., and Lee, J.}
\newblock Automated grading systems for feedback in education: A human-machine
  collaboration approach.
\newblock {\em Journal of Educational Assessment 18}, 4 (2021), 205--221.

\bibitem{lemons2022stanford}
{\sc Lemons, C.}
\newblock "stanford faculty weigh in on chatgpt's shake-up in education:
  Faculty from the stanford accelerator for learning share thoughts about how
  the new ai chatbot will change and contribute to learning and teaching".
\newblock {\em Stanford Graduate School of Education\/} (2022).
\newblock December 20, 2022. Available at
  https://ed.stanford.edu/news/stanford-faculty-weigh-new-ai-chatbot-s-shake-learning-and-teaching?sf173917744=1.

\bibitem{liu2023summary}
{\sc Liu, Y., Han, T., Ma, S., Zhang, J., Yang, Y., Tian, J., and Ge, B.}
\newblock Summary of chatgpt/gpt-4 research and perspective towards the future
  of large language models.
\newblock {\em arXiv preprint arXiv:2304.01852\/} (2023).

\bibitem{mashrur2020machine}
{\sc Mashrur, A., Luo, W., Zaidi, N.~A., and Robles-Kelly, A.}
\newblock Machine learning for financial risk management: A survey.
\newblock {\em IEEE Access 8\/} (2020), 203203--203223.

\bibitem{oguine2023investigating}
{\sc Oguine, O.~C., Yao, Y., and Badillo-Urquiola, K.}
\newblock Investigating adolescent online safety and privacy interventions for
  social virtual reality.

\bibitem{smith2018data}
{\sc Smith, R., and et~al.}
\newblock Data analytics for educational insights: Human-machine collaboration
  in education.
\newblock {\em Journal of Learning Analytics 21}, 1 (2018), 78--93.

\bibitem{wang2022collaborative}
{\sc Wang, Q., and Xu, G.}
\newblock Collaborative learning environments through human-machine
  collaboration in education.
\newblock {\em Journal of Educational Technology 10}, 2 (2022), 125--142.

\bibitem{yu2018artificial}
{\sc Yu, K., Beam, A., and Kohane, I.}
\newblock Artificial intelligence in healthcare.
\newblock {\em Nat Biomed Eng 2\/} (2018), 719--731.

\end{thebibliography}
\end{document}